
\documentclass[twocolumn]{aastex631}
\usepackage[]{graphicx}

\shortauthors{Clarkson et al.}

\begin{document}

\title{First Frequency-Time-Resolved Imaging Spectroscopy Observations of Solar Radio Spikes}

\author[0000-0003-1967-5078]{Daniel L. Clarkson}
\affiliation{School of Physics \& Astronomy, University of Glasgow, Glasgow, G12 8QQ, UK}

\author[0000-0002-8078-0902]{Eduard P. Kontar}
\affiliation{School of Physics \& Astronomy, University of Glasgow, Glasgow, G12 8QQ, UK}

\author[0000-0003-2291-4922]{Mykola Gordovskyy}
\affiliation{Department of Physics \& Astronomy, University of Manchester, Manchester M13 9PL, UK}

\author[0000-0002-4389-5540]{Nicolina Chrysaphi}
\affiliation{LESIA, Observatoire de Paris, Universit\'{e} PSL, CNRS, Sorbonne Universit\'{e}, Universit\'{e} de Paris, 5 place Jules Janssen, 92195 Meudon, France}
\affiliation{School of Physics \& Astronomy, University of Glasgow, Glasgow, G12 8QQ, UK}

\author[0000-0002-6872-3630]{Nicole Vilmer}
\affiliation{LESIA, Observatoire de Paris, Universit\'{e} PSL, CNRS, Sorbonne Universit\'{e}, Universit\'{e} de Paris, 5 place Jules Janssen, 92195 Meudon, France}
\affiliation{Station de Radioastronomie de Nan\c{c}ay, Observatoire de Paris, CNRS, PSL, Universit\'{e} d'Orl\'{e}ans, Nan\c{c}ay, France}


\received{2021 June 30}
\revised{2021 July 26}
\accepted{2021 Aug 4}
\submitjournal{ApJL}

\begin{abstract}
    Solar radio spikes are short duration and narrow bandwidth fine structures in dynamic spectra observed from GHz to tens of MHz range. Their very short duration and narrow frequency bandwidth are indicative of sub-second small-scale energy release in the solar corona, yet their origin is not understood. Using the LOw Frequency ARray (LOFAR), we present spatially, frequency and time resolved observations of individual radio spikes associated with a coronal mass ejection (CME). Individual radio spike imaging demonstrates that the observed area is increasing in time and the centroid positions of the individual spikes move superluminally parallel to the solar limb. Comparison of spike characteristics with that of individual Type IIIb striae observed in the same event show similarities in duration, bandwidth, drift rate, polarization and observed area, as well the spike and striae motion in the image plane suggesting fundamental plasma emission with the spike emission region on the order of ${\sim}\:10^8$~cm, with brightness temperature as high as $10^{13}$~K. The observed spatial, spectral, and temporal properties of the individual spike bursts are also suggesting the radiation responsible for spikes escapes through anisotropic density turbulence in closed loop structures with scattering preferentially along the guiding magnetic field oriented parallel to the limb in the scattering region. The dominance of scattering on the observed time profile suggests the energy release time is likely to be shorter than what is often assumed. The observations also imply that the density turbulence anisotropy along closed magnetic field lines is higher than along open field lines.  
\end{abstract}

\section{Introduction}

Solar activity sporadically releases magnetic energy via solar flares and coronal mass ejections (CMEs) that brightly manifest via electromagnetic radiation from X-rays to radio waves \citep[e.g.][as a review]{2011SSRv..159..107H}. Solar radio bursts are a signature of electrons accelerated in flares and CMEs. Solar radio spikes are short duration ($10-1000$~ms) bursts with narrow spectral widths from $\Delta{f}/f\simeq 0.002-0.01$, observed from $7-8$~GHz \citep{1986SoPh..104..117S,1992A&AS...93..539B} down to decametric frequencies \citep{2014SoPh..289.1701M}. Spike durations are observed to decrease with increasing frequency to below $10$~ms at gigahertz frequencies \citep{1986SoPh..104...99B, 1986SoPh..104..117S}. Their short duration and narrow frequency range are indicative of processes that occur on millisecond timescales and hence provide a unique avenue to study the fastest processes in the solar corona \citep[e.g.][]{2002SSRv..101....1A,2021ApJ...910..108K}. Indeed, electron acceleration due to magnetic energy release leads to formation of electron beams that subsequently excite Langmuir waves that produce the observed radio emission. Therefore, spike durations represent an upper limit for the energy release time. Spikes are observed in dynamic spectra either chaotically or grouped in clusters and in connection with either Type III \citep{1972A&A....17..267T, 1990A&A...231..202G, 2016SoPh..291..211S, 2017pre8.conf..381M}, Type II \citep{1984SoPh...92..329K, 2019A&A...624A..76A}, or Type IV \citep{1967ApJ...147..711M, 1990A&A...231..202G, 2016SoPh..291..211S, 2016A&A...586A..29B} solar radio bursts. Spikes are most abundant between ${\sim}\:300-3000$~MHz \citep{1986SoPh..104...99B}. Their frequency drift rates are reported to vary from zero to $>100$~MHz s$^{-1}$ \citep{1972A&A....16...21T}, while their fluxes tend not to exceed a few hundred solar flux units (sfu; 1 sfu = $10^{-22}$~W m$^{-2}$ Hz$^{-1}$). At decametric wavelengths, spikes show time profiles similar to that of Type III radio bursts but with shorter durations \citep{2014SoPh..289.1701M}.

It was suggested that the plasma emission mechanism thought to produce Type III bursts \citep{1958SvA.....2..653G} is responsible for spike emission \citep{1975A&A....39..107Z}. A comparison of spike emission with Type III bursts has been made by \cite{1972A&A....17..267T} who notes that the exciter spatial extent determines whether a spike or Type III burst is produced. \cite{1972A&A....16...21T} proposed that weak electron beams with lower densities and smaller spatial sizes than those that produce Type III bursts are responsible for spike bursts. Another proposed mechanism is electron cyclotron maser (ECM) emission \citep{1980IAUS...86..457H} due to spikes being observed in conjunction with Type IV radio bursts, and could serve to explain spike emission in regions of strong magnetic fields and/or low densities. \cite{2011ApJ...743..145C} suggest ECM emission as the source of a spike burst related to a powerful X-class flare, but notes that below 130 MHz, the densities required would be too high in a post-eruption loop system for ECM to operate. Moreover, if one of the brightest events recorded cannot produce sufficient conditions for ECM emission high in the corona, then it is infeasible for less powerful events.

Imaging of spike clusters has mainly been conducted at higher frequencies: \cite{2006A&A...457..319K} image radio spikes at $327$ and $410.5$~MHz using the Nan\c{c}ay Radio Heliograph (NRH) \citep{10.1007/BFb0106458} finding the spike emission region above the soft and hard X-ray sources of the associated flare, but find no significant motion of the spike sources. The radio images show the spike bursts were temporally and spatially associated with compressed magnetic field structures due to a CME. \citet{2002A&A...383..678B} also show spike locations away from the flare site with two cases near flare loop tops. Interestingly, the spikes are observed during the flare decay phase, indicating a possible link to post-flare acceleration sites. Imaging by \citet{1995A&A...302..551K} shows the spike emission at  altitudes where it is proposed that the energy release causing both spike emission and Type III emission occurs. More recent VLA imaging in the $1.0-1.6\;\mathrm{GHz}$ range show the spike source to be located above the flare arcade \citep{2021ApJ...911....4L}. However, the detailed spatial evolution of individual spikes in time, space and frequency have not been reported in the literature before.

In this paper, we report for the first time, the frequency and time-resolved evolution of individual radio spikes produced before and in the wake of a CME using the LOw Frequency ARray \citep[LOFAR;][]{2013A&A...556A...2V} tracking of the spike source motion, as well as the spike characteristics in dynamic spectra between $30-70$~MHz. The spike observations reveal superluminal source motion and source size expansion at 100 ms scales consistent with strong anisotropic scattering of radio-waves in a turbulent corona. The spike locations before and after the CME are shifted upwards, presumably perturbed by the CME. The results also confirm not only the similarity of the observed spike properties with Type IIIb striae, but the co-spatial character of the Type IIIb and spike sources.

\section{Spike characteristics}

\begin{figure*}[t!]
    \centering
    \epsscale{1.2}
    \plotone{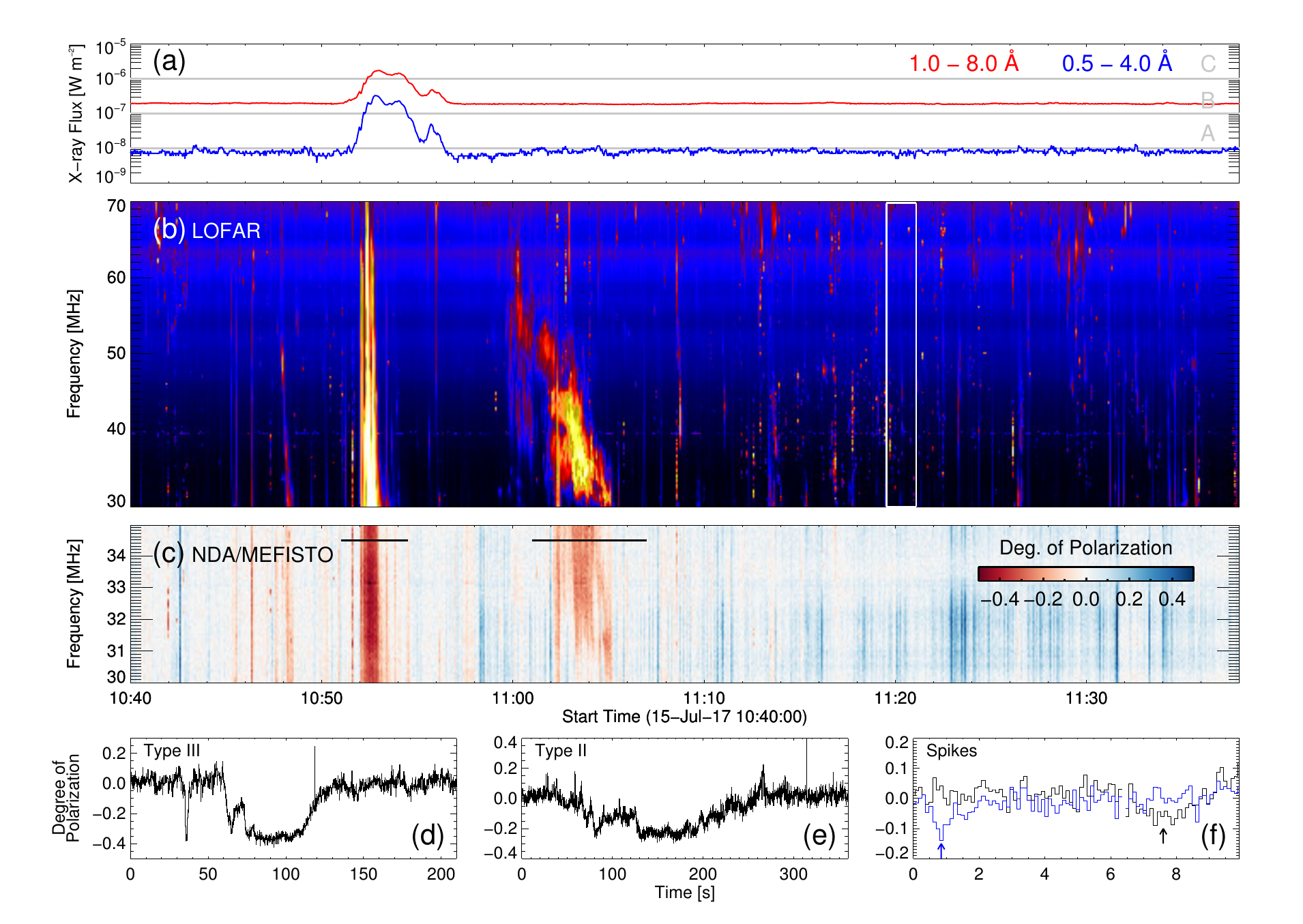}
    \caption{X-ray and radio emissions between 10:40 and 11:40 UT. \textbf{(a)} X-ray lightcurves from the GOES spacecraft \citep{1994SoPh..154..275G}. \textbf{(b)} Dynamic spectrum of the LOFAR observation. A series of bright Type III bursts can be seen near 10:52, with a Type II burst near 11:03. Spikes are distributed across the dynamic spectrum with a cluster between 11:10 and 11:21. The white box represents the region shown in Figure \ref{fig:ds_overview}(a). \textbf{(c)} Dynamic spectrum of the circular polarization from the NDA/MEFISTO receiver. 
    \textbf{(d-f)} Polarization from the NDA/MEFISTO receiver of the Type III and Type II bursts, respectively, at 34.5 MHz indicated by the solid black lines in panel (c), and polarization of two spike bursts within the region shown in Figure \ref{fig:ds_overview}(c) at 32.50 and 33.88 MHz. The spikes are highlighted by the arrows.}
    \label{fig:ds_1hr}
\end{figure*}

\begin{figure*}
    \centering
    \epsscale{1.2}
    \plotone{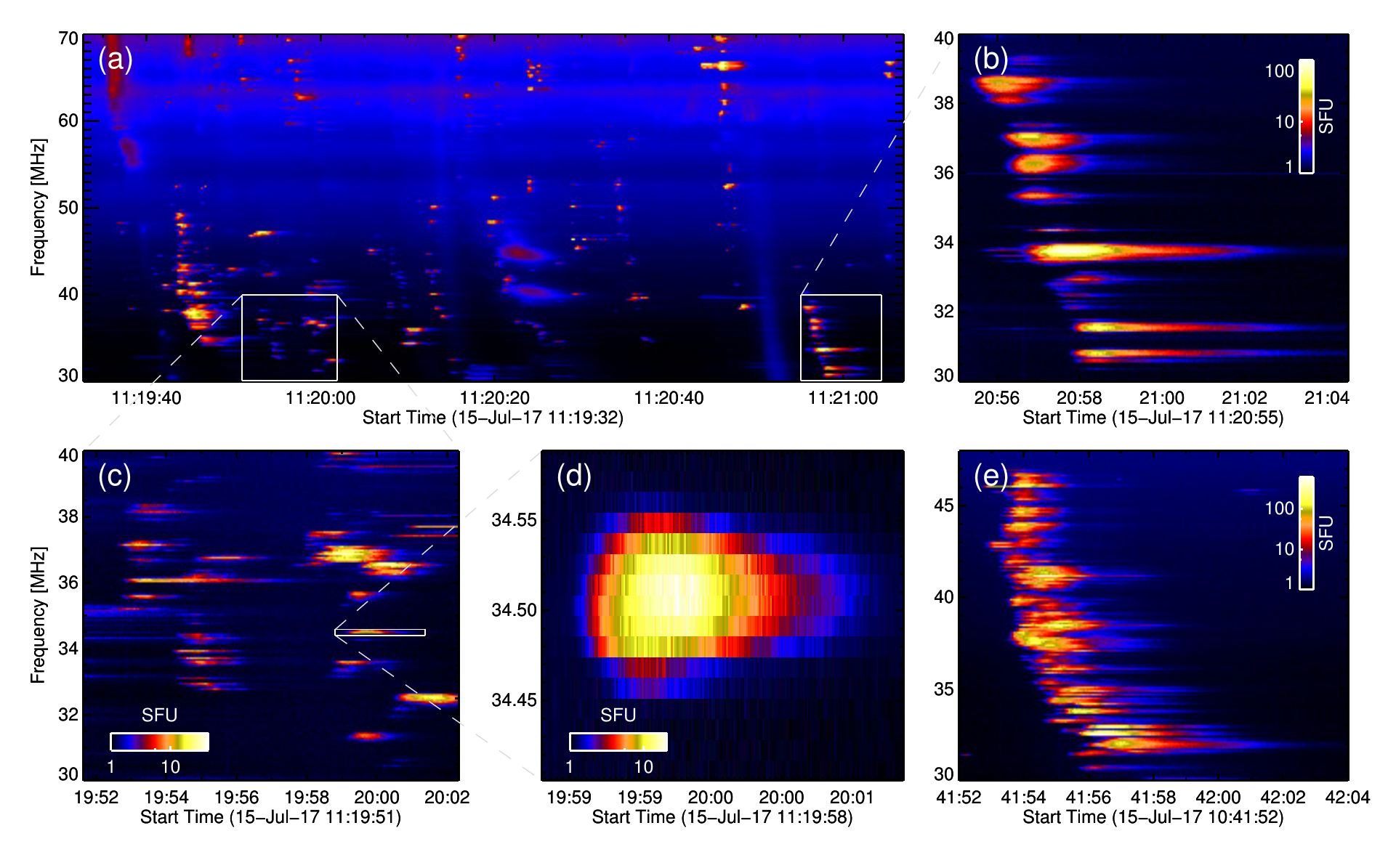}
    \caption{Zoomed in dynamic spectra of numerous spike and Type IIIb features within Figure \ref{fig:ds_1hr}. \textbf{(a)} Numerous spikes between 11:19:32 and 11:21:06 UT from the region bounded by the white box in Figure \ref{fig:ds_1hr}. \textbf{(b)} Type IIIb burst near 11:21 UT. \textbf{(c)} Spike cluster between $30-40$~MHz. \textbf{(d)} An individual radio spike. \textbf{(e)} Type IIIb burst observed near 10:42 UT.}
    \label{fig:ds_overview}
\end{figure*}

The active region AR12665 on the western solar limb produced a C1.4 class solar flare between 10:50 to 10:55, with the ejection of a bifurcated jet \citep{2020ApJ...893..115C}. Solar soft X-ray flux, radio flux and polarization dynamic spectra during the eruptive event are shown in Figure \ref{fig:ds_1hr}. The polarization measurements are provided by the Nan\c{c}ay Decameter Array (NDA) MEFISTO receiver \citep{1980Icar...43..399B, 2000GMS...119..321L, 2013Lecacheux, 10.1553/PRE8s455}, designed to automatically filter noise. Numerous short duration and low frequency bandwidth spikes are seen. Some spikes are chaotically distributed in the spectrum, whilst others form chains similar to Type IIIb bursts. A Type II burst observed by LOFAR near 11:03 UT is associated with the CME, as reported by \citet{2020ApJ...893..115C}. In addition, a cluster of bright Type III bursts occur close to the start of the jet eruption near 10:52 UT, along with two Type IIIb bursts at 10:42 and 11:21 UT shown in Figure \ref{fig:ds_overview}(b,e).

43 isolated solar radio spikes between 10:40 to 11:36 UT are analysed using LOFAR tied-array beam-forming mode \citep{2013A&A...556A...2V} using 24 core Low Band Antenna stations in the outer LBA configuration with a maximum baseline of $3.6\;\mathrm{km}$ in the frequency interval $30-70$~MHz. 216 interferometrically synthesized beams image the solar corona up to $\sim 3$~R$_{\odot}$ with a temporal and spectral resolution of $10$~ms and $12.2$~kHz, respectively. This enabled spikes with durations $<1$~s and spectral widths greater than ${\sim}\:24$~kHz to be individually resolved. The spike observations were temporally decreased in resolution to 20 ms to reduce noise. The flux was calibrated using observations of Taurus A \citep[see][for details]{2017NatCo...8.1515K}. Figure \ref{fig:ds_overview} shows zoomed-in versions of the fine structures.

\begin{figure*}
    \minipage{0.33\textwidth}
      \includegraphics[width=\linewidth]{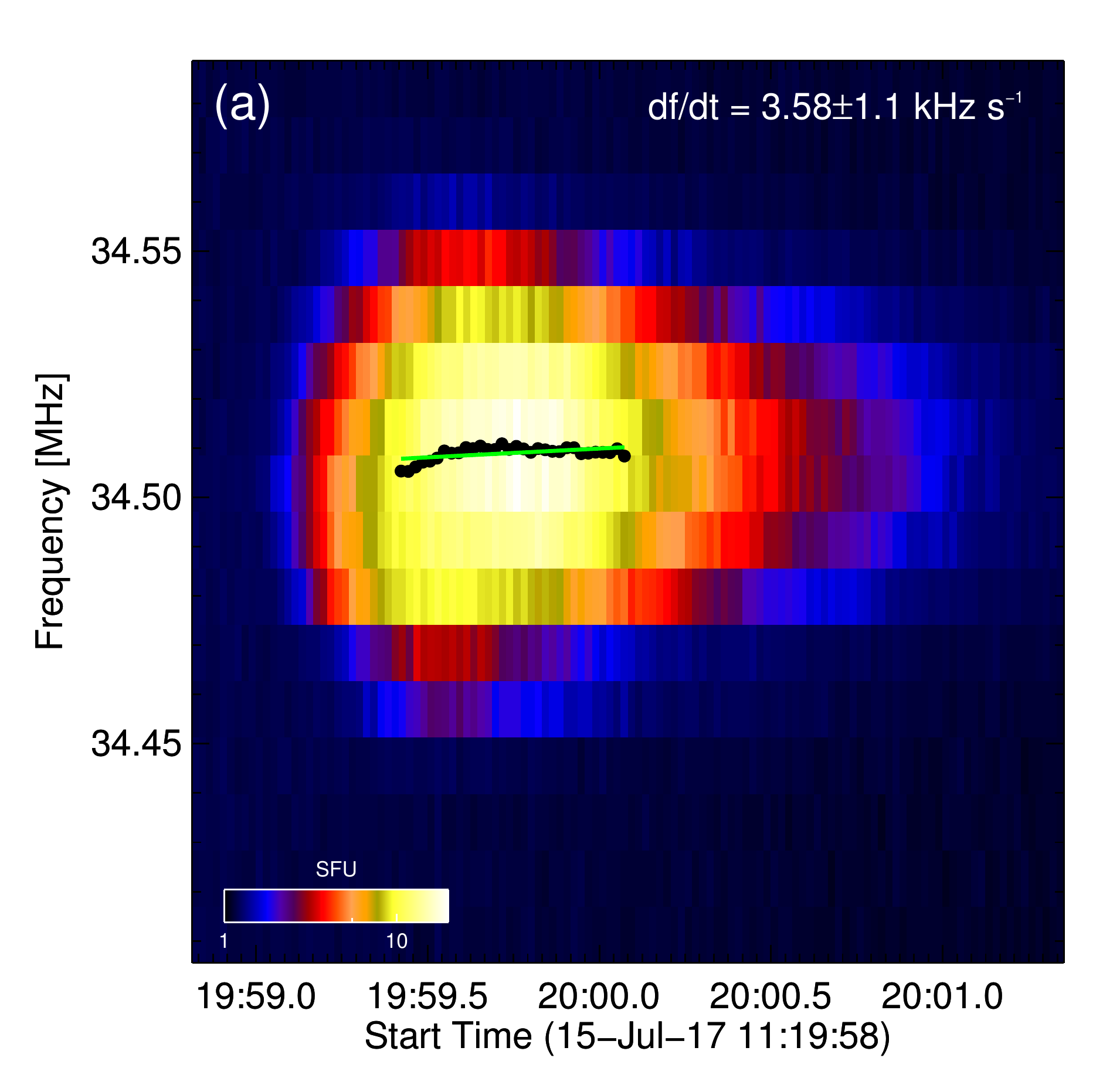}
    \endminipage\hfill
    \minipage{0.33\textwidth}
      \includegraphics[width=\linewidth]{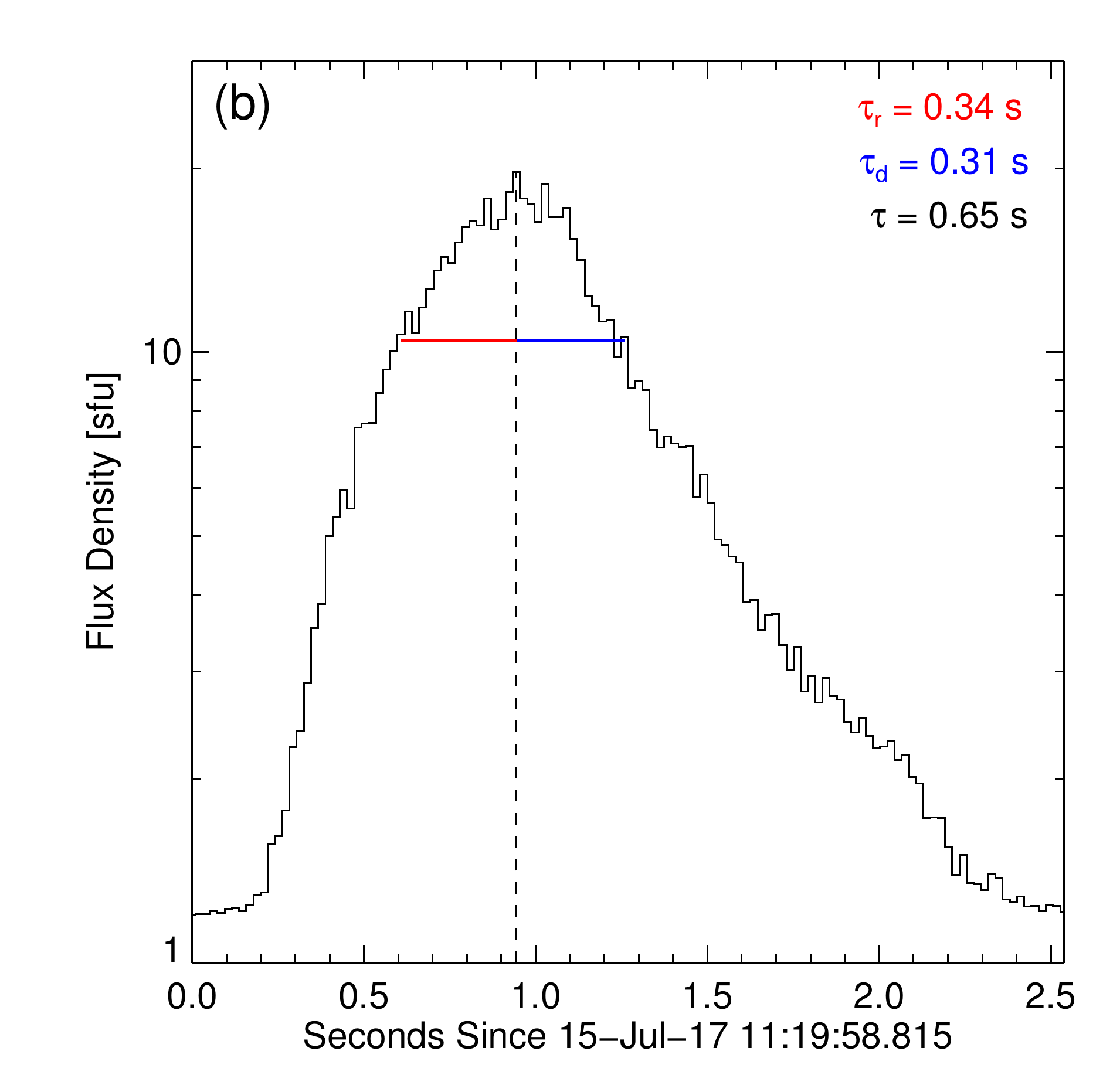}
    \endminipage\hfill
    \minipage{0.33\textwidth}%
      \includegraphics[width=\linewidth]{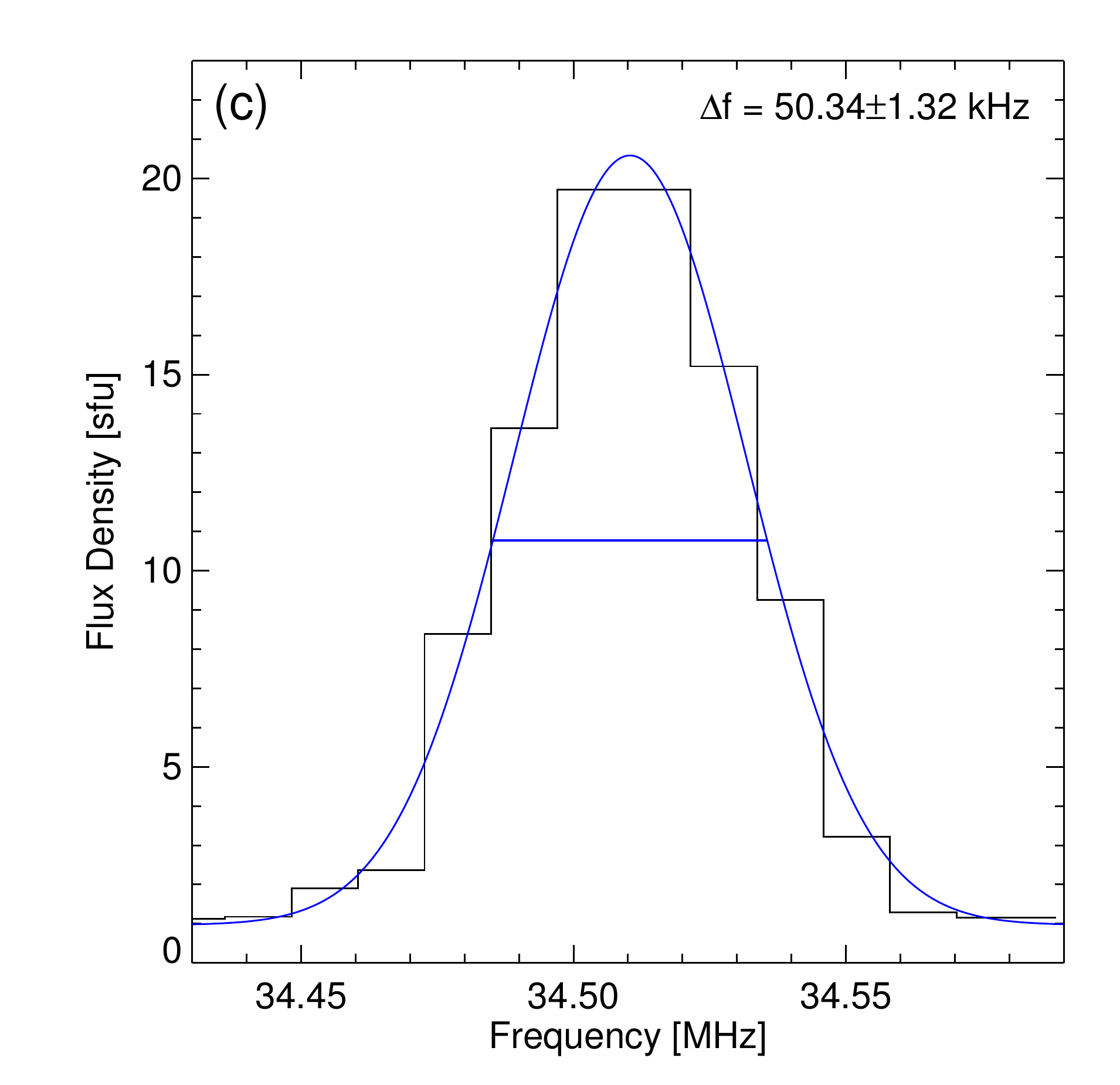}
    \endminipage\hfill
    \minipage{0.33\textwidth}
      \includegraphics[width=\linewidth]{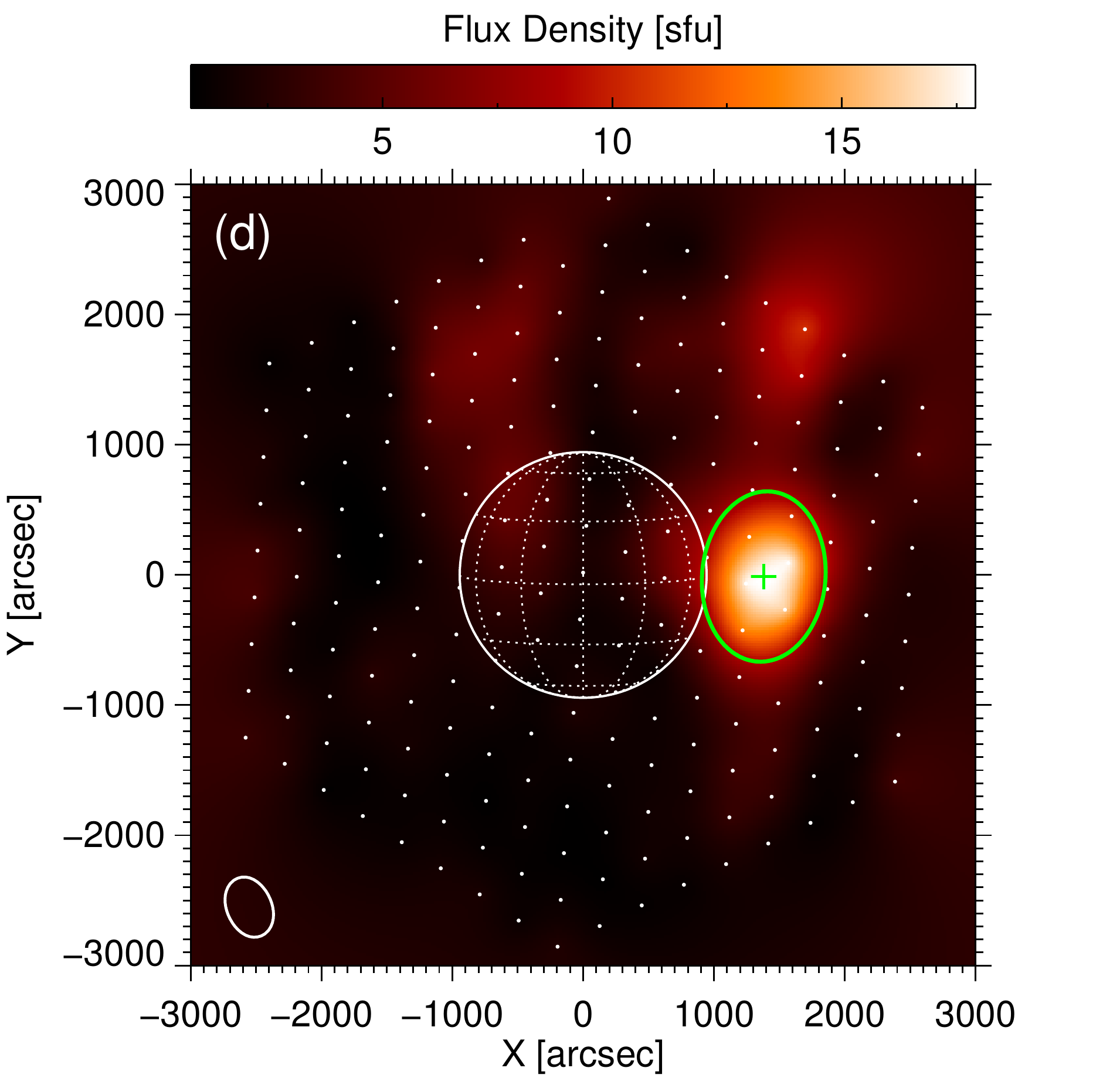}
    \endminipage\hfill
    \minipage{0.33\textwidth}
      \includegraphics[width=\linewidth]{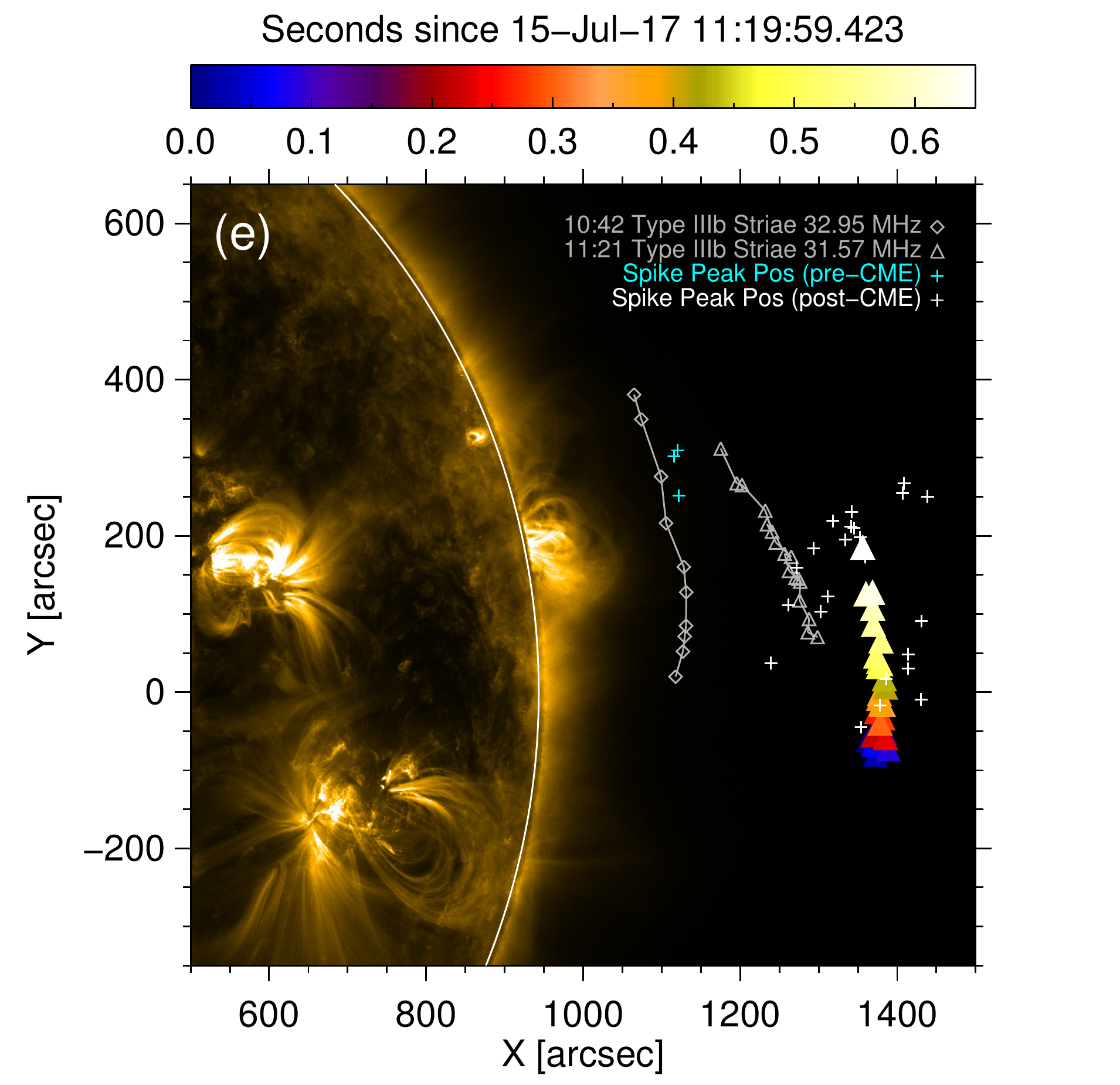}
    \endminipage\hfill
    \minipage{0.33\textwidth}%
      \includegraphics[width=\linewidth]{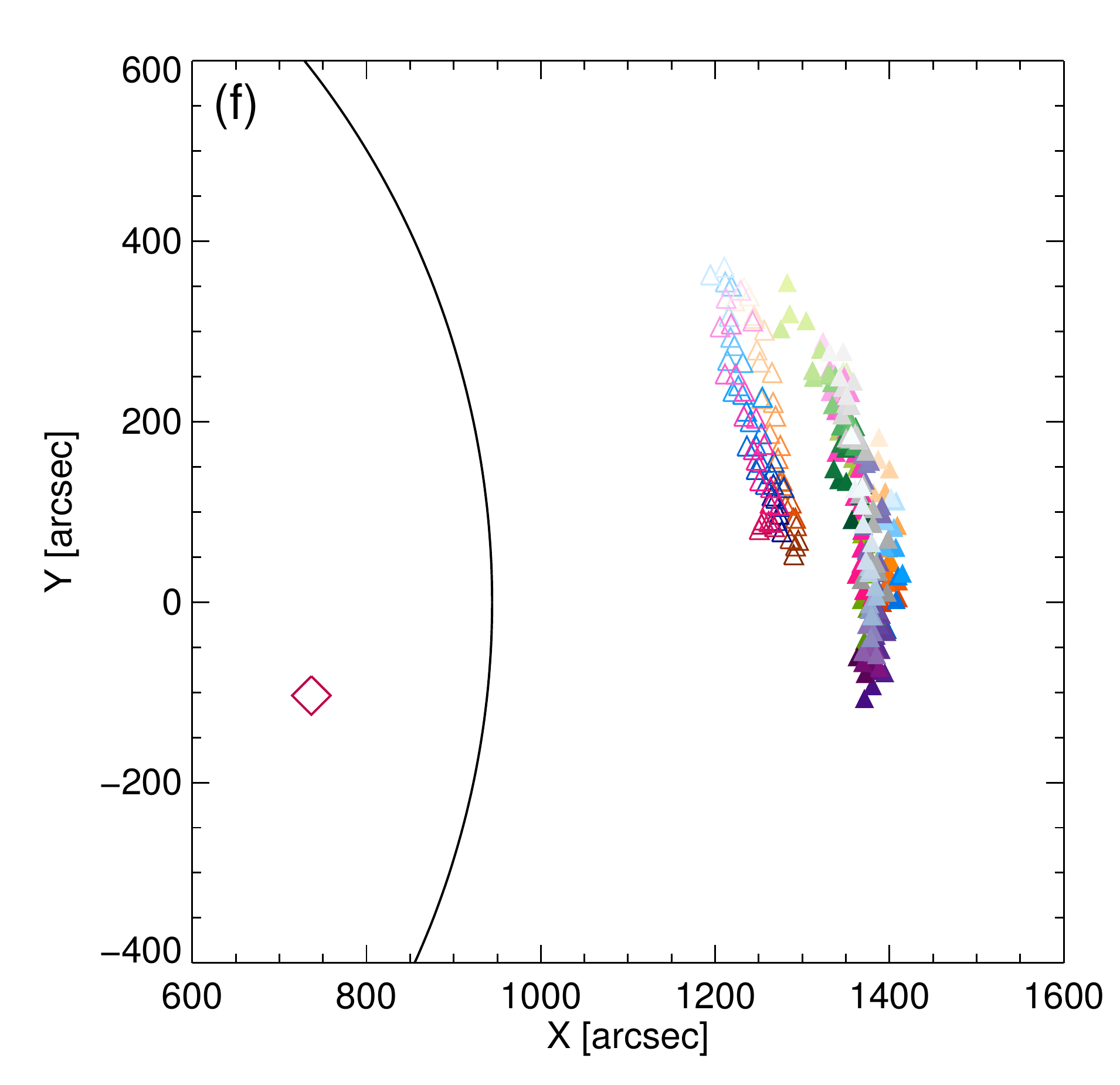}
    \endminipage
    \caption{\textbf{(a)} Dynamic spectrum of the single spike shown in Figure \ref{fig:ds_overview}(d). 
    The black points show the peak of the Gaussian fit to the flux profile 
    at each time bin. The green line shows the linear fit to these points to derive the frequency drift rate. 
    \textbf{(b)} Spike flux profile at $34.5$~MHz. 
    The dashed vertical line shows the burst peak time, 
    with the horizontal red and blue lines showing the rise and decay time, respectively. \textbf{(c)} Flux profile at the time of the burst peak (black) with a Gaussian fit (blue). The horizontal line marks the FWHM spectral width.
    \textbf{(d)} LOFAR image at the peak intensity of the spike at $34.5$~MHz near 11:19:59.8 UT. The green contour marks the 2D Gaussian fit at the FWHM level with the centroid marked by the green cross. 
    The white dots represent the phased-array beam pointing, 
    and the white oval shows the half-maximum synthesized LOFAR beam. 
    \textbf{(e)} Spike centroid motion in time at a fixed central spike frequency of $34.5$~MHz represented by the colored triangles overlaid on an SDO/AIA 171 \r{A} image at 11:20:57 UT. 
    The blue plus symbols show the peak centroid position of the spikes between $30-45$~MHz before the CME, 
    whilst the white plus symbols show the spike peak centroid positions after the CME. 
    The grey lines with diamond markers represent the centroid motion of a single Type IIIb striae near 10:42 UT at $32.95$~MHz, whereas the open grey triangles represent a striae near 11:21 UT at $31.57$~MHz. \textbf{(f)} Motion of 10 individual spikes. 
    The open triangles represent frequencies between $43-45$ MHz and closed triangles between $33-35$~MHz. The color gradients represent time increasing from dark to light. The red diamond shows the approximate location of the active region.}
    \label{fig:spike_properties}
\end{figure*}

\begin{figure*}
    \minipage{0.33\textwidth}
      \includegraphics[width=\linewidth]{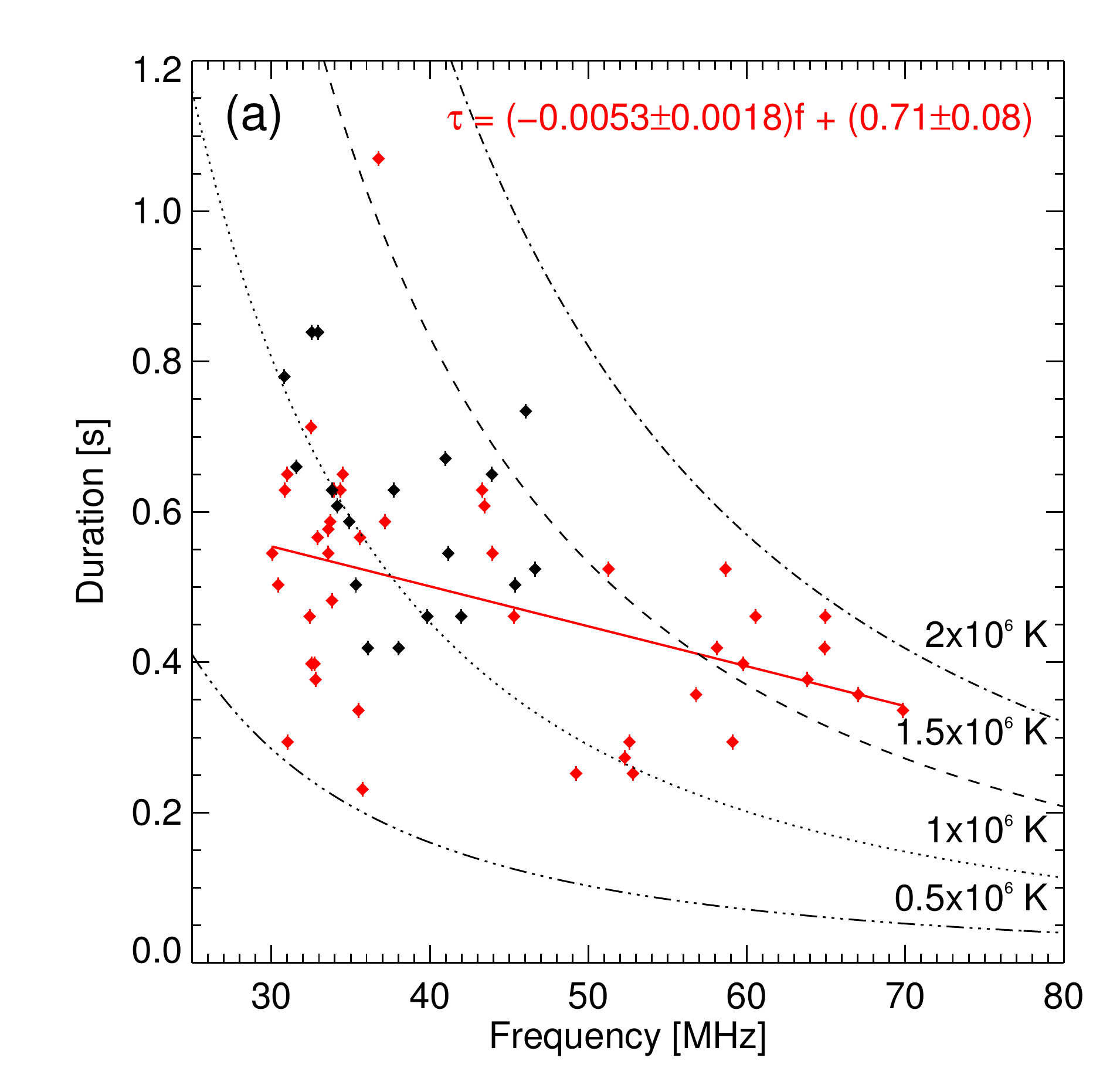}
    \endminipage\hfill
    \minipage{0.33\textwidth}
      \includegraphics[width=\linewidth]{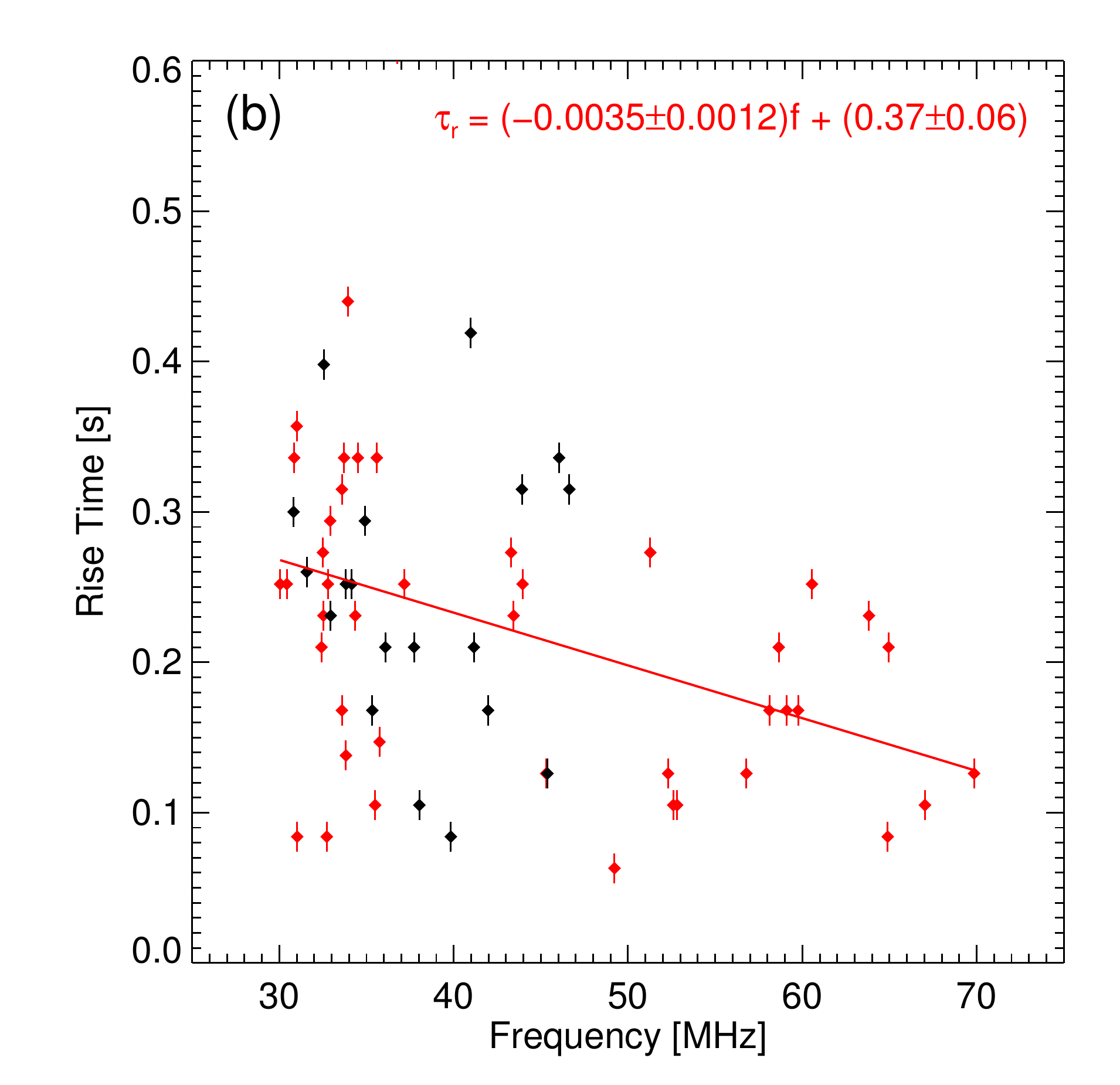}
    \endminipage\hfill
    \minipage{0.33\textwidth}%
      \includegraphics[width=\linewidth]{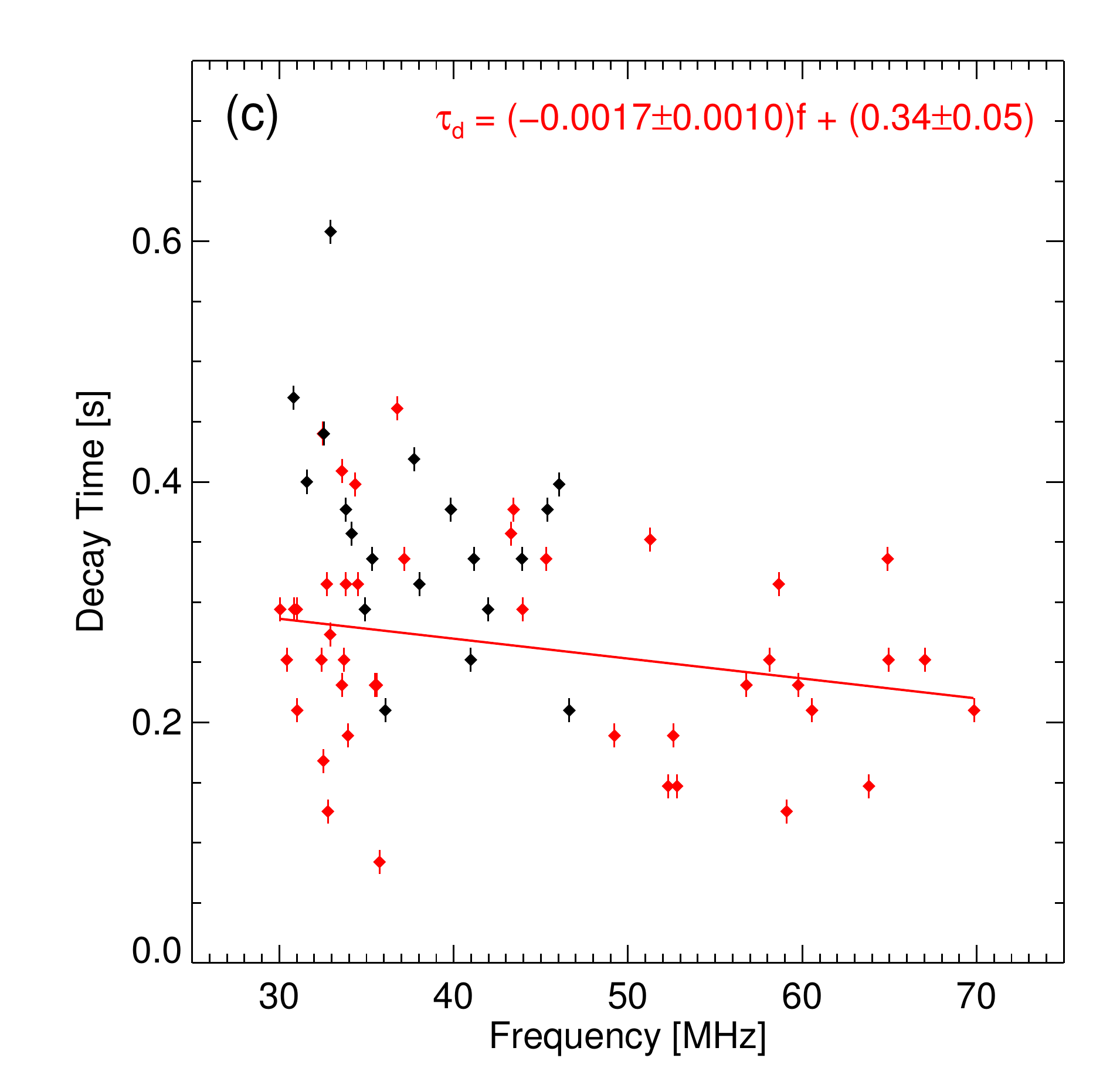}
    \endminipage\hfill
      \minipage{0.33\textwidth}
      \includegraphics[width=\linewidth]{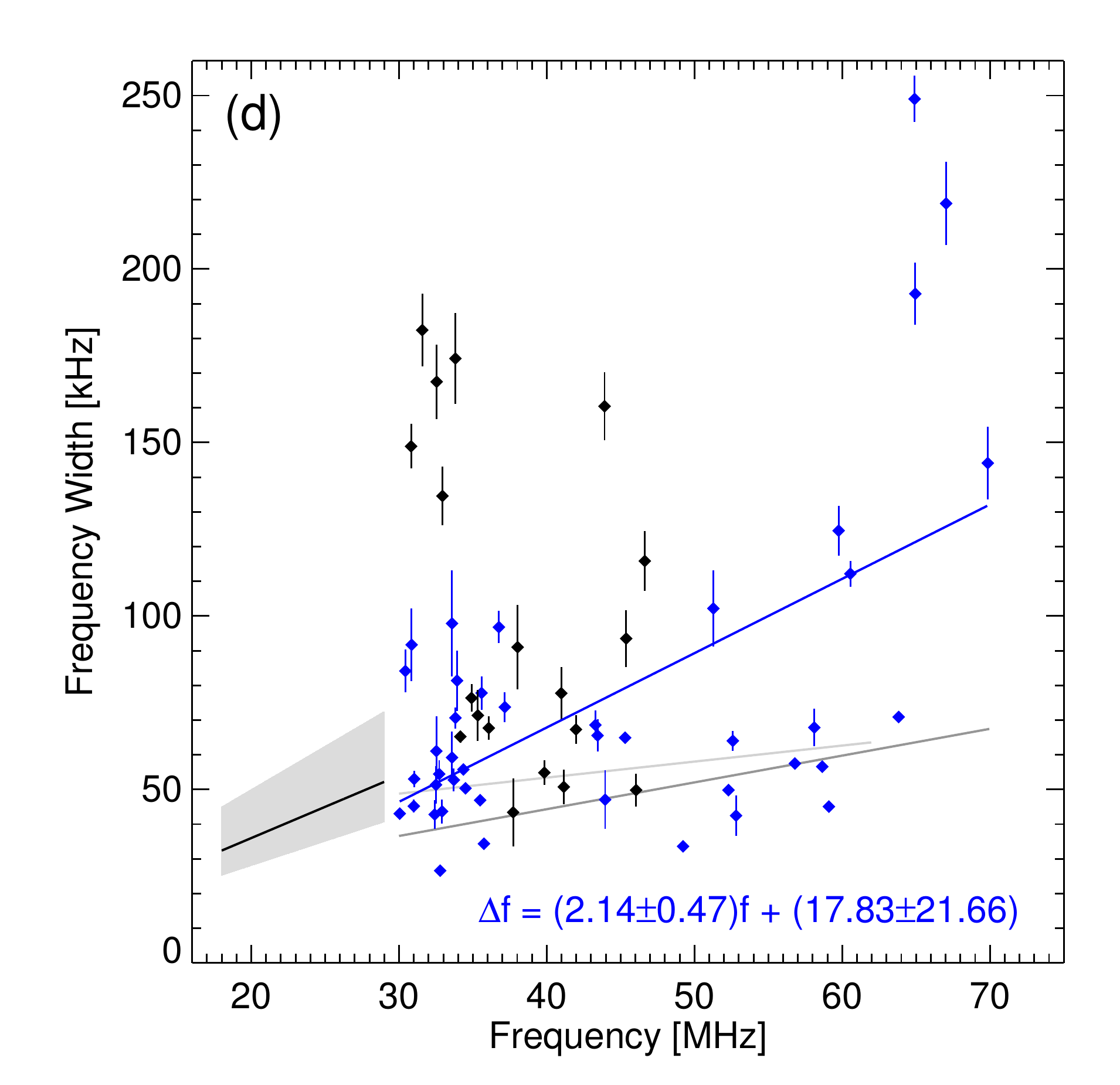}
    \endminipage\hfill
    \minipage{0.33\textwidth}
      \includegraphics[width=\linewidth]{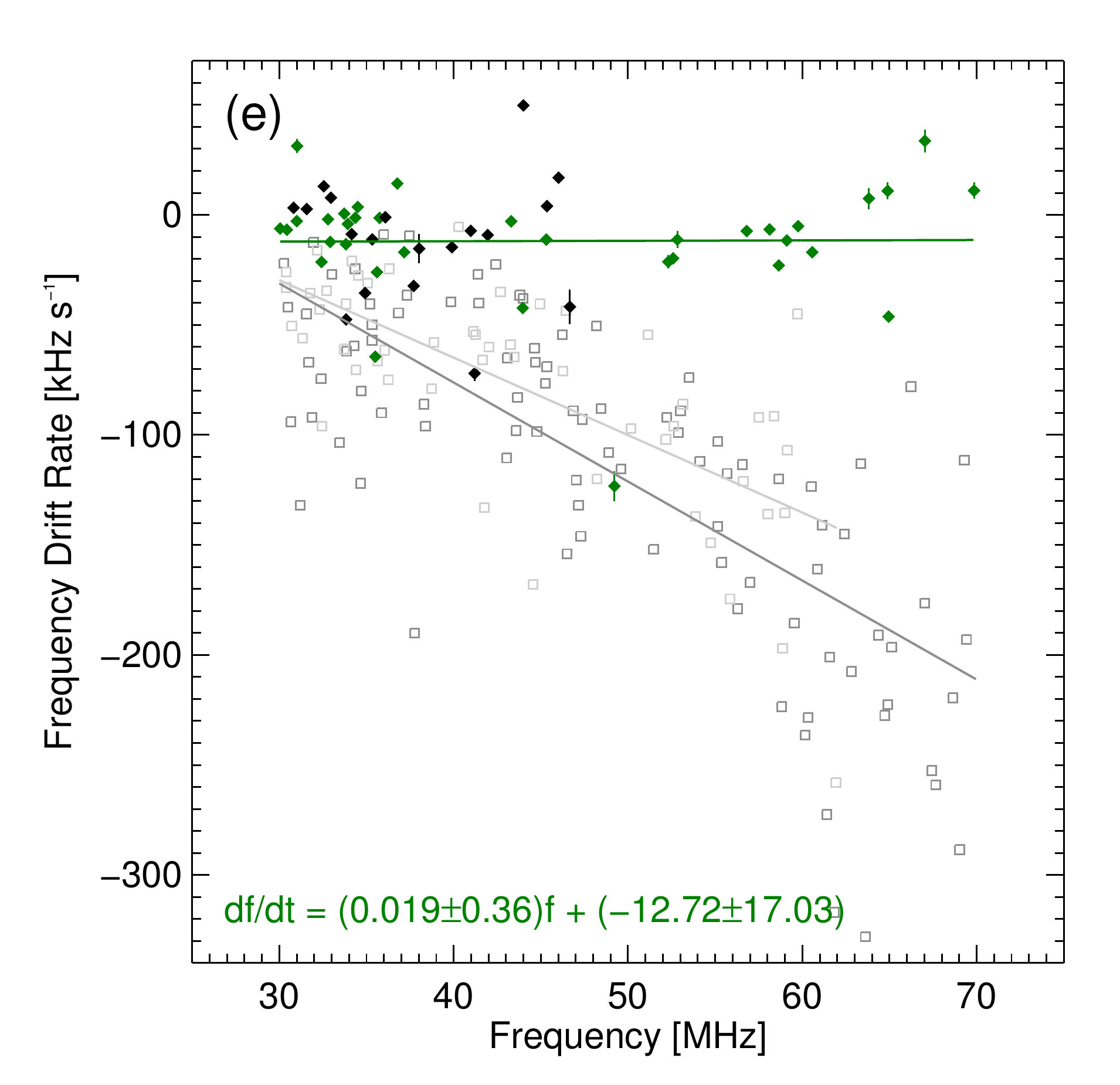}
    \endminipage\hfill
    \minipage{0.33\textwidth}%
      \includegraphics[width=\linewidth]{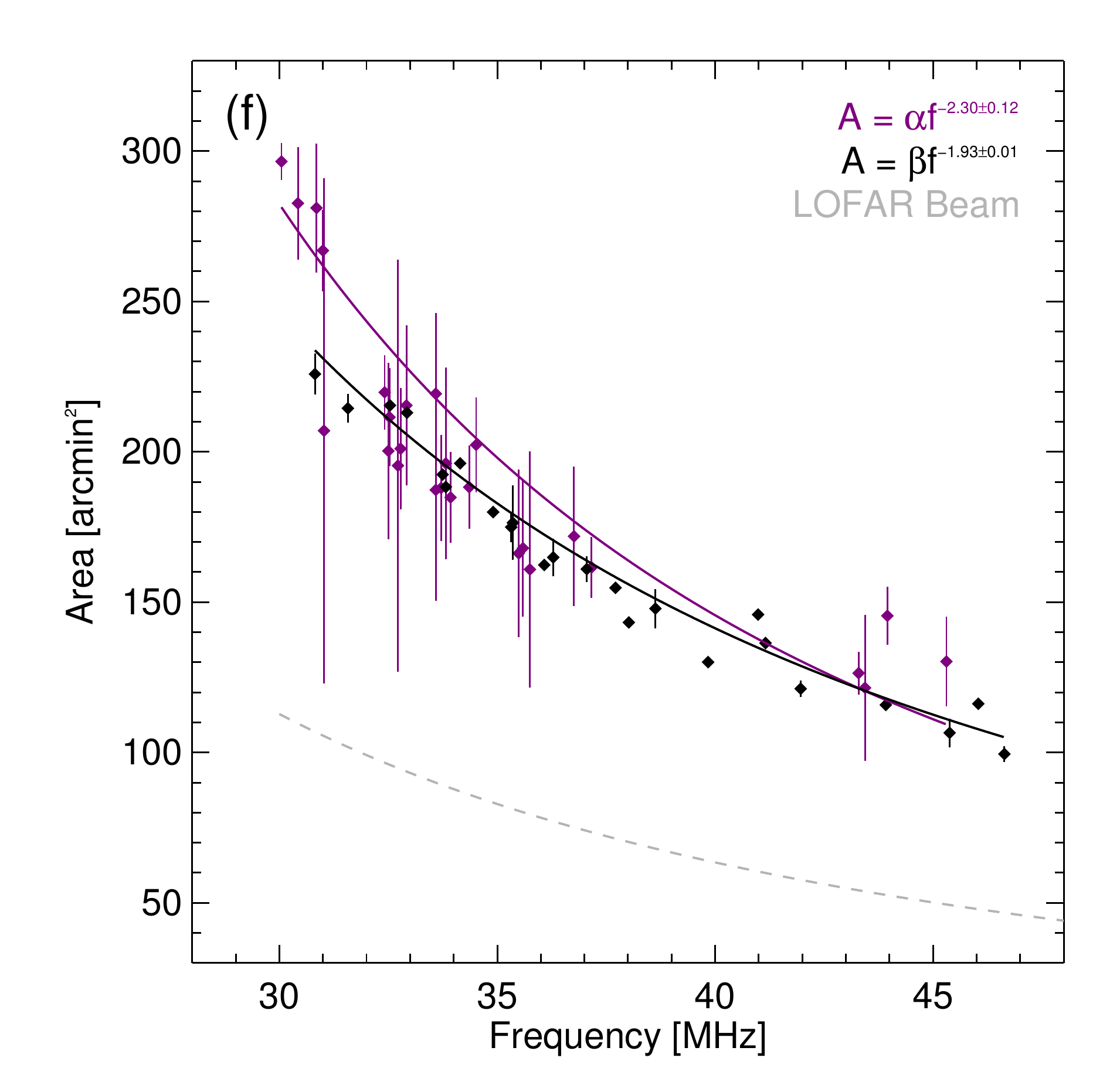}
    \endminipage
    \caption{Spike characteristics from LOFAR data between $30-70$~MHz. 
    The black diamonds represent striae from Type IIIb bursts during the same event. \textbf{(a)} Spike FWHM durations (red). The thin-line curves show the collision time,
     $\tau_{\mathrm{coll}}\approx T^{3/2}/110n$, \citep{1967ApJ...147..711M} for a plasma of the temperatures $T$, where $n$ is the plasma density. 
     The red line shows the linear fit to the spike durations. 
     \textbf{(b)} Spike FWHM rise times with linear fit. 
     \textbf{(c)} Spike FWHM decay times with linear fit. The errors in (a-c) represent the uncertainty due to the temporal resolution. 
     \textbf{(d)} Spike spectral widths with a linear fit (blue). The uncertainty represents the 1-sigma error provided by the Gaussian fitting procedure. The black solid line shows the average fit to spike spectral widths from \cite{2014SoPh..289.1701M} with the grey region showing the upper and lower bounds. The two grey lines show fits to striae widths from \cite{2018SoPh..293..115S}. \textbf{(e)} Frequency drift rates of spikes between $30-70$~MHz with a linear fit (green). 
     The uncertainty represents the 1-sigma error provided by the Gaussian fitting procedure. 
     The light and dark grey squares show the striae drift rates from \cite{2018SoPh..293..115S} with corresponding linear fits. 
     \textbf{(f)} Observed FWHM area of spikes at their peak intensity (purple). 
     The coefficients $\alpha$ and $\beta$ of the powerlaw fits are $(7.09\pm3.0)\times10^5$ and $(1.75\pm0.085)\times10^5$, respectively. 
     The light-grey dashed line shows the LOFAR beam area. The errors are calculated as in \cite{2017NatCo...8.1515K}.}
    \label{fig:spike_characteristics}
\end{figure*}

The spike peak fluxes range from $4-66$~sfu, averaging at ${\sim}\:18$~sfu. The peak flux of the Type IIIb striae are an order of magnitude brighter than the average spikes, peaking up to ${\sim}\:200$~sfu. The time profiles of solar radio spikes at a given frequency resembles that of Type III bursts; a prompt rise time followed by a longer decay. However, the duration of spikes near $30$~MHz are shorter up to a factor of ${\sim}\:20$. Figure \ref{fig:spike_properties}(b) shows the time series at the central spike frequency of $34.5$~MHz with a FWHM duration of $0.65$~s. Typically, the rise time is shorter than the decay time, averaging at $0.22$ and $0.26$~s, respectively. The spike durations fall between $0.2-1.1$~s, average at $0.48$~s (Figure \ref{fig:spike_characteristics}(a)), with a rate of change of $-5\pm1.8$~ms per MHz. The durations are comparable with nearby Type IIIb striae as well as the collision time for a plasma of temperatures between $(0.5-2)\times10^6$~K shown by the curves in Figure \ref{fig:spike_characteristics}(a).

The instantaneous frequency flux profile at the burst peak time is symmetrical (Figure \ref{fig:spike_properties}(c)). The Gaussian fit gives a FWHM bandwidth of $\Delta{f}=50.34\pm1.32$~kHz (Figure \ref{fig:spike_properties}(c)). Average bandwidth is $76.1$~kHz between $30-70$~MHz and tends to increase with frequency (Figure \ref{fig:spike_characteristics}(d)), with some spike widths reaching up to $250$~kHz near $70$~MHz. Spike spectral widths overlap with that of Type IIIb striae between $30-46$~MHz from the same event, with striae widths also observed up to $174$~kHz.

The spike in Figure \ref{fig:spike_properties}(a) shows a near-zero drift rate of $\mathrm{d}f/\mathrm{d}t=3.58\pm1.1$~kHz/s. The majority of spikes have negative drift rates between zero and $-70$~kHz/s with an outlier near $-123$~kHz/s and eight spikes that have a positive drift rate (Figure \ref{fig:spike_characteristics}(e)). The linear fit indicates no significant change in frequency drift with frequency. The spike drift rates are comparable with striae drift rates ranging between $50$ to $-70$~kHz/s. In comparison, the bulk Type IIIb structures shown in Figure \ref{fig:ds_overview}(b,e) have drift rates of $-3.14\pm0.21$ and $-2.31\pm0.26$~MHz/s, respectively, measured by a linear fit to the peak flux position at the central frequency of each striae.

\begin{figure*}[!htb]
    \minipage{0.33\textwidth}
      \includegraphics[width=\linewidth]{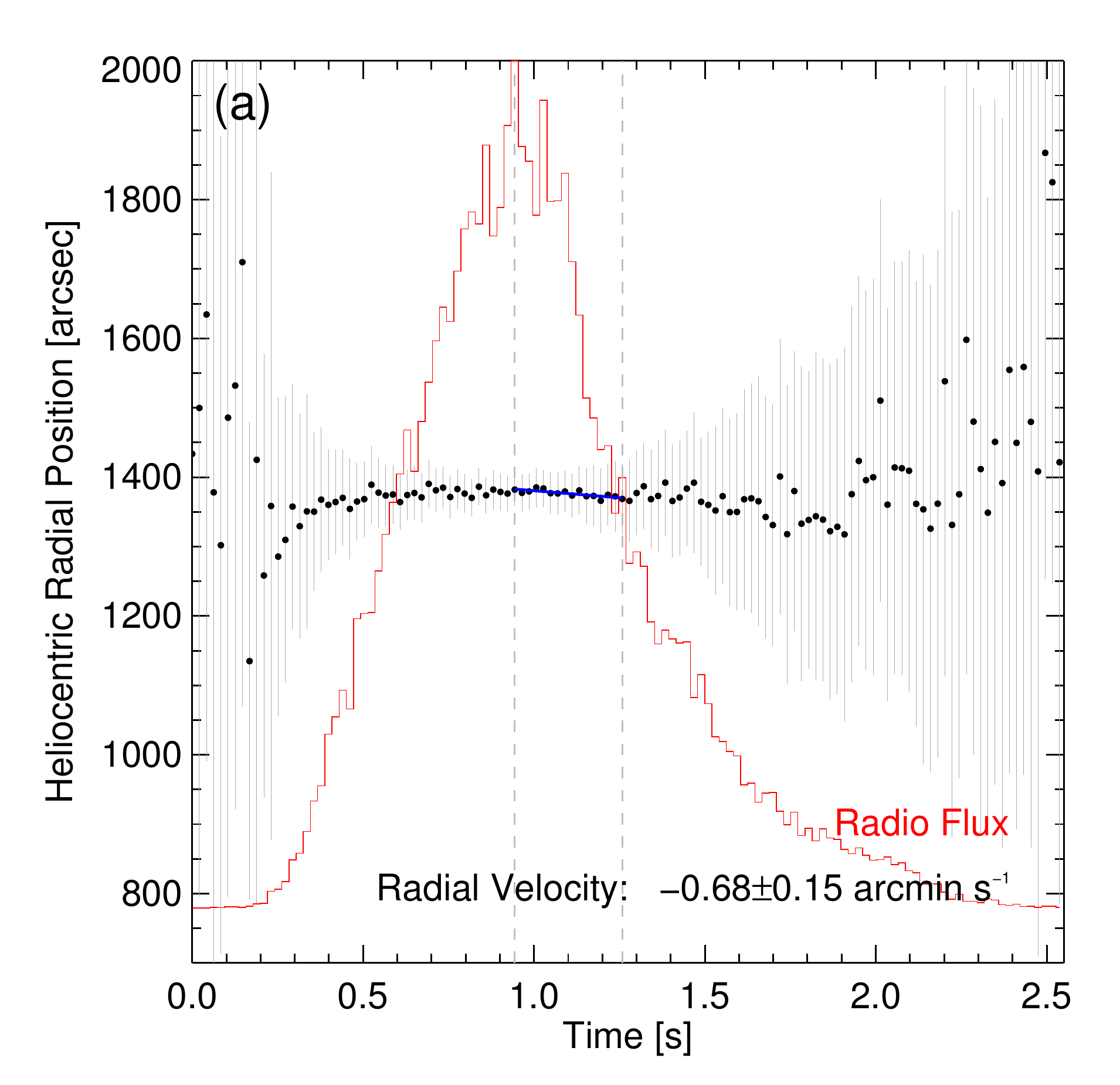}
    \endminipage\hfill
    \minipage{0.33\textwidth}
      \includegraphics[width=\linewidth]{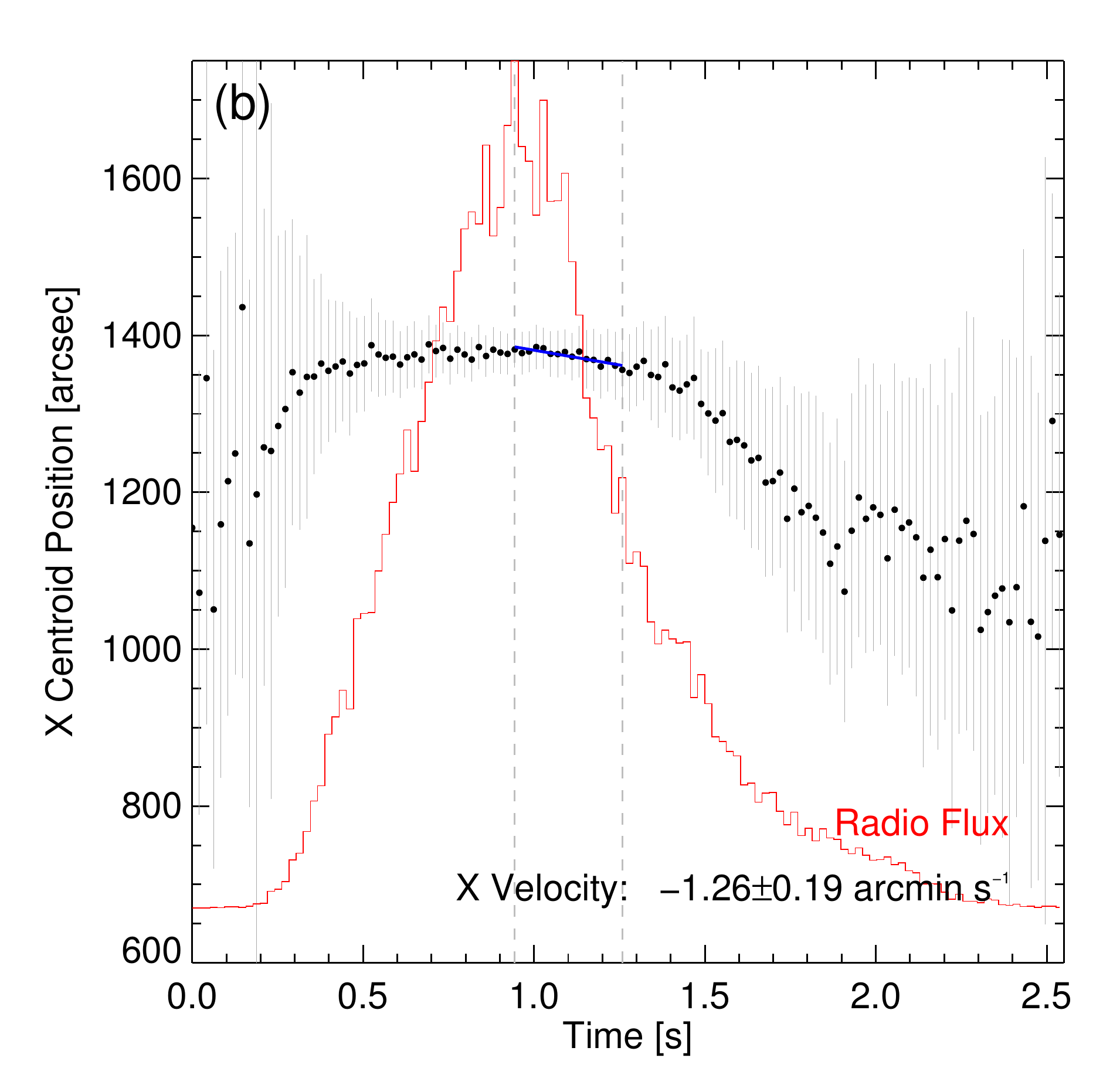}
    \endminipage\hfill
    \minipage{0.33\textwidth}%
      \includegraphics[width=\linewidth]{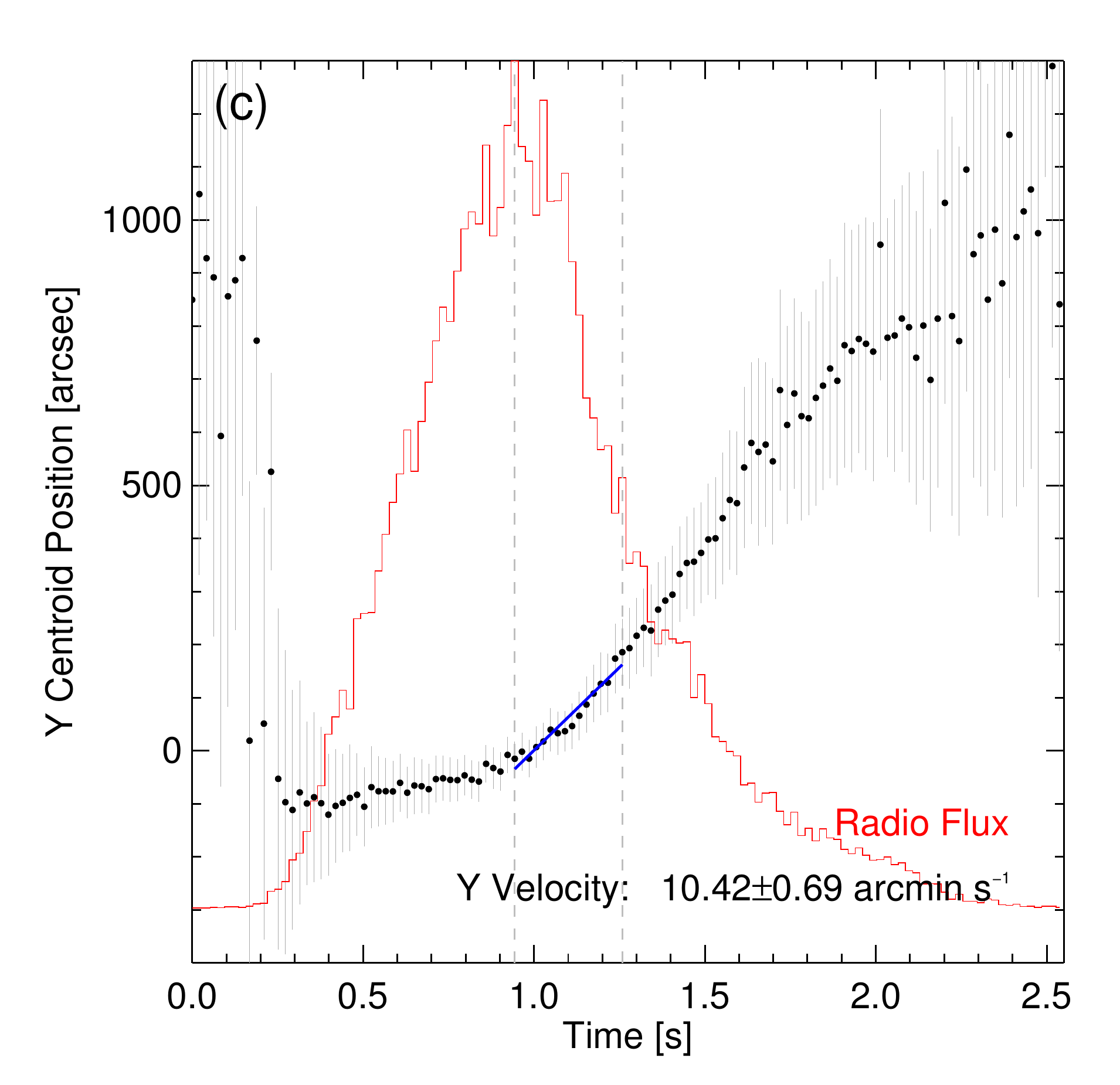}
    \endminipage\hfill
    \minipage{0.33\textwidth}
      \includegraphics[width=\linewidth]{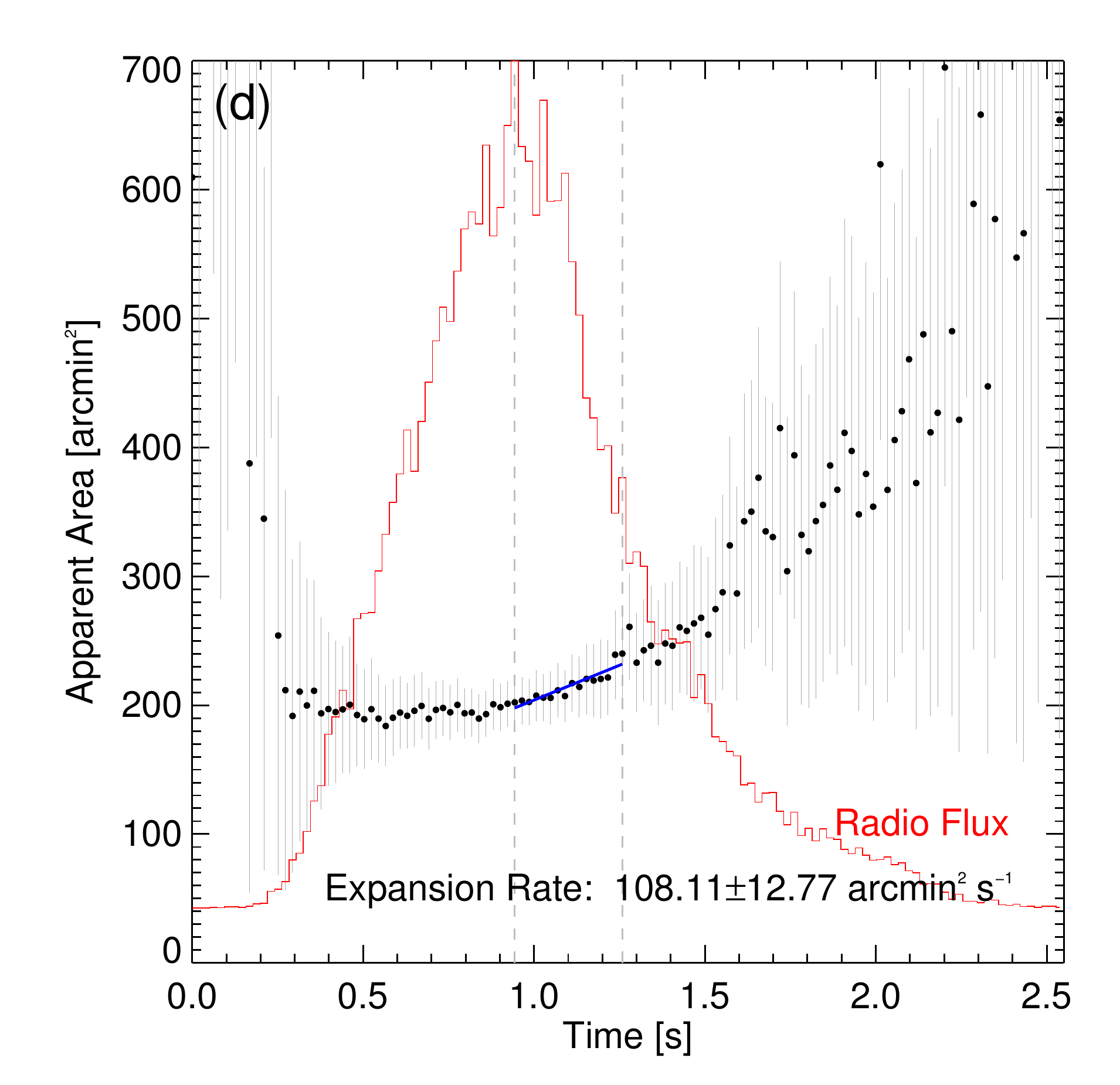}
    \endminipage\hfill
    \minipage{0.33\textwidth}
      \includegraphics[width=\linewidth]{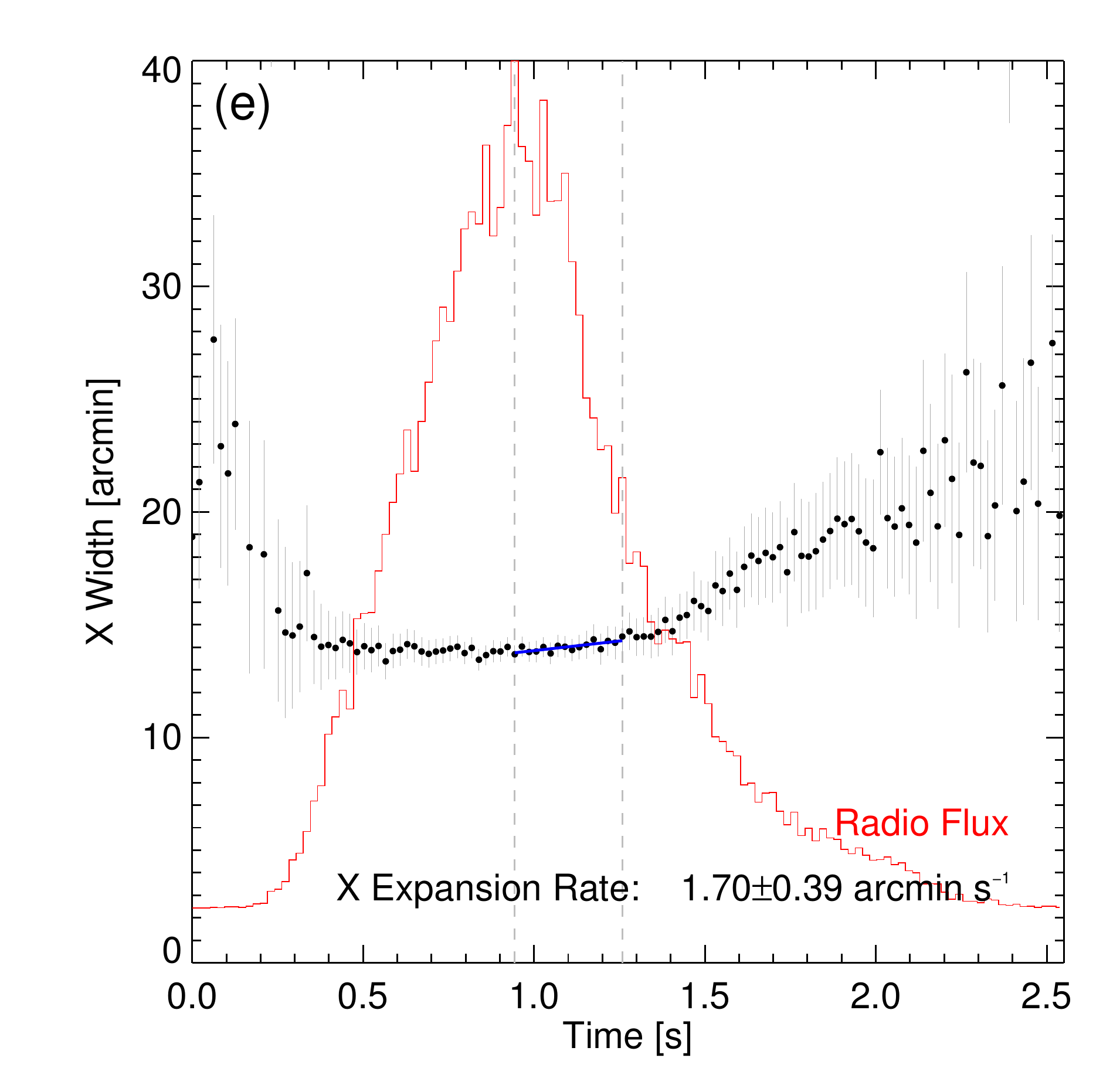}
    \endminipage\hfill
    \minipage{0.33\textwidth}%
      \includegraphics[width=\linewidth]{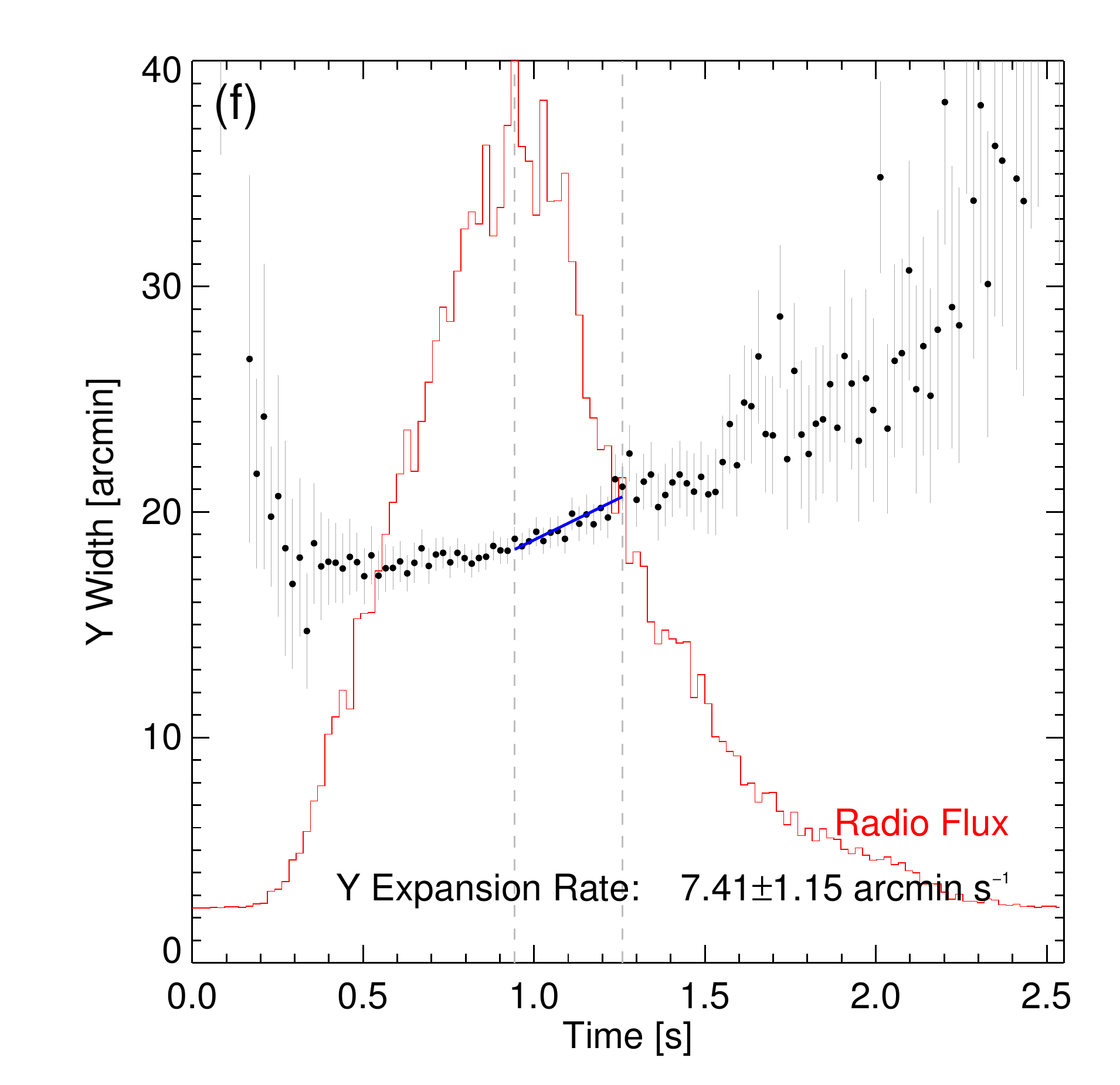}
    \endminipage

    \caption{Spike source motion and expansion over the duration of the dynamic spectrum shown in Figure \ref{fig:ds_overview}(d) at $34.5$~MHz. The red curve shows normalised radio flux at the same frequency, with the vertical grey dashed lines at the peak and FWHM decay time. The blue lines show linear fits over the decay period. The errors are given as in \cite{2017NatCo...8.1515K}. \textbf{(a)} Heliocentric radial position. \textbf{(b)} Spike centroid velocity in the X direction. \textbf{(c)} Spike centroid velocity in the Y direction. \textbf{(d)} Observed area and \textbf{(e,f)} widths in the X and Y direction over time.}
    \label{fig:radial_xc_yc_motion_area}
\end{figure*}

The observed spike FWHM area at the peak of the central frequency is $202.19\pm16.3$~arcmin$^2$ at $34.5$~MHz (Figure \ref{fig:spike_properties}(d)). The observed spike areas decrease with increasing frequency from $297$ to $122$~arcmin$^2$ between $30$ to $45$~MHz (Figure \ref{fig:spike_characteristics}(f)) in a similar manner to drift-pair bursts \citep{2019A&A...631L...7K}, and is approximated with a power law $A\sim f^{-\gamma}$ where $\gamma=2.3$ and $1.9$ for spikes and striae, respectively. The large uncertainties are due to their low intensities. The synthesised LOFAR beam area over this frequency range is $A_\mathrm{beam}\simeq 50-113$~arcmin$^2$, so the LOFAR-beam-corrected source areas are up to $A\simeq 72-184$~arcmin$^2$. Importantly, the spike areas at a fixed frequency increase over time at tens of ms scales (Figure \ref{fig:radial_xc_yc_motion_area}(d)) with expansion rates between $18-108$~arcmin$^2$/s that are most pronounced during the decay phase, similar to Type IIIb observations \citep{2017NatCo...8.1515K}.

One of the intriguing observations is the variability of the radio spike sources (positions and areas) with time at tens of millisecond scales. The spike centroid position moves vertically in the plane-of-sky image over $0.65$~s across the solar equator (Figure \ref{fig:spike_properties}(d)), covering  $262$~arcsec $\simeq 0.28$~R$_{\odot}$ in the image plane, corresponding to the speed of light over the FWHM duration. The frequency drift rate inferred velocity, assuming a Newkirk density model \citep{1961ApJ...133..983N}, is ${\sim}\:45.1$~km/s. Spikes with higher frequency drifts such as $60$~kHz/s at $35.5$~MHz show speeds of up to $680$~km/s. The spikes observed ${\sim}\:10$~mins before the flare-CME appear closer to the disk centre, while the spikes in the wake of the CME are further away (Figure \ref{fig:spike_properties}(e)). The spike source motion follows a trajectory parallel to the Type IIIb striae observed prior to the CME. Post-CME Type IIIb striae show motion that originates within the same region as the post-CME spikes (Figure \ref{fig:spike_properties}(f)), suggesting a common exciter. It is interesting to note that the FWHM spike source areas are comparable to Type IIIb areas in this event, but smaller than observed before \citep[e.g.][]{2017NatCo...8.1515K}.

Figure \ref{fig:radial_xc_yc_motion_area}(a-c) shows the spike centroid motion in the image plane for the radial, \textit{x}, and \textit{y} positions. The radial distance changes weakly, with the bulk motion in the \textit{y}-direction with a superluminal plane-of-sky speed of $10.4$~arcmin/s $\simeq 1.5c$ during the decay phase where the velocity is most pronounced. This vertical motion is typical for all observed spikes during this period (see Figure \ref{fig:spike_properties}(f)), with \textit{x} and \textit{y} drift rates between $-0.44$ to $-2.27$~arcmin/s and $5.25$ to $11.18$~arcmin/s, respectively. In the \textit{y}-direction, this corresponds to apparent speeds between $0.76c-1.8c$.

\section{Summary}

The spectral and temporal characteristics of the observed spikes are consistent with those previously reported at similar frequencies \citep{2014SoPh..289.1701M,2016SoPh..291..211S}. There is a tendency for shorter durations and higher bandwidths towards higher frequencies; a similar trend is observed near GHz frequencies \citep{1990A&A...231..202G,2016A&A...586A..29B,1993A&A...274..487C,1992A&AS...93..539B}. The magnetic field strength at this coronal height ${\sim}\:1.78$~R$_{\odot}$ by \citet{1978SoPh...57..279D}, $B(r)=0.5(r/R_\odot - 1)^{-1.5}$ ~G, is $0.73$~G, which is close to estimates using LOFAR Type IIIb observations by \citet{2018ApJ...861...33K}. The electron cyclotron frequency for this field strength is $f_{ce}\simeq 2.0$~MHz $\ll f_{pe}\sim 30$~MHz, so the criteria for ECM emission is not satisfied. Moreover, NDA/MEFISTO measurements indicate that the Type III and Type II bursts both show left-handed circular polarization up to $-0.4$ and $-0.3$, respectively (Figure \ref{fig:ds_1hr}). The spikes within the region shown in Figure \ref{fig:ds_overview}(c) are also left-hand polarized at $-0.15$ and $-0.1$, indicating that the spikes are produced by the same emission mechanism. Therefore, we suggest plasma emission as the source of the radio spikes similar to Type IIIb bursts, but likely from weaker/slower electron beams. The plasma turbulence also changes the spatial distribution of Langmuir waves (seen in the numerical simulations by \citet{2001A&A...375..629K}), creating regions of enhanced Langmuir waves and hence electromagnetic emission seen as striae or spikes.

The radio spike positions are observed before and after the onset of the solar eruptive event at 10:50 UT. The location of radio spikes and Type IIIb striae after  the CME are $1250-1450$~arcsec away from the Sun centre, compared to ${\sim}\:1100$~arcsec prior to the CME eruption showing a shift away from the Sun. For the first time, the time-resolved observations of individual spikes reveal source motions and source size changes at 100~ms scales. The spike sources (both before and after the CME) follow trajectories approximately parallel to the solar limb, which contrasts the previously observed radial motions of Type IIIb bursts \citep{2017NatCo...8.1515K, 2019ApJ...885..140Z}, and the corresponding spherically symmetric coronal simulations \citep{2020ApJ...905...43C}.

The motion of the spike sources is superluminal ($0.76c-1.8c$) and accompanied by the superluminal FWHM source size expansion of $7.4$~arcmin/s ${\sim}\:1.1c$. For scatter-dominated sources, the source velocity depends on the angle between the line-of-sight and the direction from the Sun centre to the source location, as well as the anisotropy of plasma turbulence \citep{2019ApJ...884..122K, 2020ApJ...905...43C, 2020ApJ...898...94K}. Large heliocentric angles such as that observed by spikes in this observation are subject to larger displacements along the direction of the guiding magnetic field and increased apparent velocities. Superluminal motions are, in fact, observed in radio-wave propagation simulations \citep[see Figure 5 by][]{2020ApJ...898...94K}, suggesting $1.0c-1.1c$ speeds for sources located at heliocentric angles of 30 and 50 degrees.

The previously unobserved non-radial superluminal motion of spike and Type IIIb sources in this event suggests a different magnetic configuration to what was simulated by \cite{2020ApJ...905...43C}. In anisotropic density turbulence that is aligned with the magnetic field, radio waves propagate preferentially along the magnetic field direction \citep{2019ApJ...884..122K}. The observed spike and Type IIIb sources are located within the region where the magnetic field is likely forming loop-like structures \citep{2020ApJ...893..115C}, so radio-wave scattering in the region with the magnetic field lines parallel to the limb could induce the observed direction of the source motion. The extended post-reconnection closed loops are likely formed within the CME wake and is the location of weak electron beam acceleration, resulting in Langmuir wave generation and subsequent spike emission. The simulations of radio-wave transport by \citet{2020ApJ...898...94K} show that stronger anisotropy leads to smaller observed peak source sizes and superluminal velocities. Thus, the spikes and Type IIIb striae source properties are consistent with the simulations with anisotropy $\alpha =0.1-0.2$ \citep{2020ApJ...898...94K}, which is higher than the anisotropy $\alpha =0.25-0.3$ required in open configurations to explain Type III burst properties \citep[e.g.][]{2019ApJ...884..122K,2020ApJ...905...43C}. Consequently, the anisotropy of density turbulence in closed loop configurations should be higher than that along open field lines, typical for Type III bursts.

Many similarities between spike and Type IIIb striae characteristics such as duration, spectral width, velocities and observed area suggest a common physical mechanism. The spike durations decrease on average with increasing frequency, with bandwidths ranging between $20-100\;\mathrm{kHz}$, covering a similar range reported in \cite{2018SoPh..293..115S} for striae. The spike and striae drift rates show little dependence on frequency, and indicate velocities of $10-50$~km/s.  Spike drift rates close to $30$~MHz overlap with the striae drift rates presented \cite{2018SoPh..293..115S}, however the comparison diverges above $40$~MHz. The drift rate inferred velocities of our observed spikes and striae are much less than the bulk speeds of Type III bursts that propagate at characteristic speeds of ${\sim}\:10^5$~km/s $\simeq c/3$ \citep{1985srph.book..289S}. The motion of Langmuir waves \citep{2021NatAs.tmp...96R}, which is believed to be responsible for Type IIIb striae drift, could be the explanation for the individual spike drift. However, since the spikes have on average shorter duration, such drift is likely to be diluted by scattering effects.

As was noted before \cite[e.g.][]{2014SoPh..289.1701M}, the spike duration is comparable to the plasma collision time. However, a spread of temperatures (from $0.5-2$ MK) for the same cluster of spikes are required to explain the characteristic decay time (Figure \ref{fig:spike_characteristics}). Radio spikes above $100$~MHz \citep{1990A&A...231..202G} would need even higher plasma temperatures---$2-4$~MK within the collisional damping hypothesis. While damping of plasma oscillations should be present, the large source sizes, the superluminal motion of spike sources, and the aforementioned morphological similarity to Type IIIb striae implies that scattering is an important factor in determining spike profiles. Therefore, we suggest that the scattering of radio waves rather than collisional damping alone determines the time profile. Spike durations are often used to constrain the shortest energy release time in flares, with the bandwidth equating to the size of the acceleration region itself \citep{1985SoPh...96..357B}. Following the Type IIIb approach \citep{2017NatCo...8.1515K,2018SoPh..293..115S}, the size of the emitting source spike region can be estimated as  $\Delta{r}\simeq2L(\Delta{f}/f)\simeq 10^8$~cm that corresponds to a subtended solid angle of ${\sim}\:10^{-2}$~arcmin$^2$. Therefore, the effect of radio-wave propagation has increased the observed source area by four orders of magnitude. This means that the brightness temperatures of the spike sources corrected for scattering could be up to $10^{12}-10^{13}$~K, well above the values $10^8-10^9$~K when radio-wave scattering is ignored. With scattering as the determining factor of the duration and higher resulting brightness temperature, the energy release responsible for electron acceleration would be much shorter and more intense than previously assumed in the literature. The characteristic emission timescale is approximately reduced by the ratio of the observed size to the emitting region, and could be two orders of magnitude shorter, i.e. tens of milliseconds instead of ${\sim}\:1$~s as observed.

\begin{acknowledgments}
DLC, EPK, and NV are thankful to Dstl for the funding through the UK-France PhD Scheme (contract DSTLX-1000106007). EPK was supported by STFC consolidated grant ST/P000533/1. MG was supported by the STFC grant ST/T00035X/1. NC thanks CNES for its financial support. The authors acknowledge the support by the international team grant (\href{http://www.issibern.ch/teams/lofar/}{http://www.issibern.ch/teams/lofar/}) from ISSI Bern, Switzerland. This paper is based (in part) on data obtained from facilities of the International LOFAR Telescope (ILT) under project code LC8\_027. LOFAR \citep{2013A&A...556A...2V} is the Low-Frequency Array designed and constructed by ASTRON. It has observing, data processing, and data storage facilities in several countries, that are owned by various parties (each with their own funding sources), and that are collectively operated by the ILT Foundation under a joint scientific policy. The ILT resources have benefited from the following recent major funding sources: CNRS-INSU, Observatoire de Paris and Universit\'{e} d'Orl\'{e}ans, France; BMBF, MIWF-NRW, MPG, Germany; Science Foundation Ireland (SFI), Department of Business, Enterprise and Innovation (DBEI), Ireland; NWO, The Netherlands; The Science and Technology Facilities Council, UK; Ministry of Science and Higher Education, Poland. The authors thank the radio astronomy station of Nan\c{c}ay / Scientific Unit of Nan\c{c}ay of the Paris Observatory (USR 704-CNRS, supported by the University of Orleans, the OSUC and the Center Region in France) for providing access to NDA observations accessible online at \href{https://www.obs-nancay.fr}{https://www.obs-nancay.fr}.
\end{acknowledgments}

\bibliographystyle{aasjournal}
\bibliography{refs}

\end{document}